\documentclass[epj]{svjour}
\usepackage{graphicx}

\begin{document}

\title{Sum rules for X-ray magnetic circular dichroism spectra in strongly correlated ferromagnets}
\titlerunning{Sum rules for X-ray magnetic circular dichroism spectra}
\author{V.Yu.~Irkhin\inst{1} \and M.I.~Katsnelson\inst{2}}

\institute{Institute of Metal Physics, 620219 Ekaterinburg, Russia
\and Institute for Molecules and Materials, Radboud University,
6525 ED Nijmegen, The Netherlands}
\date{Received: date / Revised version: date}

\abstract{It is proven that the sum rules for X-ray magnetic
dichroism (XMCD) spectra that are used to separate spin and
orbital contributions to the magnetic moment are formally correct
for an arbitrary strength of electron-electron interactions.
However, their practical application for strongly correlated
systems can become complicated due to the spectral density weight
spreading over a broad energy interval. Relevance of incoherent
spectral density for the XMCD sum rules is illustrated by a simple
model of a ferromagnet with orbital degrees of freedom.
}
\maketitle

X-ray magnetic circular dichroism (XMCD) \cite{book,ebert} is a
powerful technique to investigate both bulk and surface magnetic
properties of materials. In particular, it allows to measure
separately spin and orbital contributions to the magnetic moment
of ferromagnets. Examples of numerous applications of this method
are recent studies of magnetism in thin metallic films \cite{Co1},
cobalt nanoparticles and clusters \cite{Co2}, in magnetite
Fe$_3$O$_4$ \cite{huang}, and in dilute magnetic alloys
\cite{imp}. A concrete way to separate the orbital and spin
magnetic moments is using the XMCD sum rules
\cite{carra,carra1,ebert}
\begin{eqnarray}
3N_{h}\frac{\int d\omega \left( \Delta \mu _{\rm L_{3}}-2\Delta
\mu _{\rm L_{2}}\right) }{\int d\omega \left( \mu _{\rm
L_{3}}^{tot}+\mu _{\rm L_{2}}^{tot}\right) } &=&\left\langle
S_{z}\right\rangle +7\left\langle
T_{z}\right\rangle \nonumber \\
2N_{h}\frac{\int d\omega \left( \Delta \mu _{\rm L_{3}}-2\Delta
\mu _{\rm L_{2}}\right) }{\int d\omega \left( \mu _{\rm
L_{3}}^{tot}+\mu _{\rm L_{2}}^{tot}\right) } &=&\left\langle
L_{z}\right\rangle \label{sumrules}
\end{eqnarray}
where $N_h$ is the number of holes in $d$-band, $\mu
_{\rm L_{2,3}}\left( \omega \right)$ are spectral intensities for
L$_{2,3}$ spectra, $\Delta \mu $ is the difference between the
spectra for left and right circularly polarized radiation and $\mu
^{tot}$ is the total absorption intensity for unpolarized one,
$S_z$, $L_z$ and $T_z$ are projections of the total spin, orbital
moment and the spin dipole operator on the magnetization
direction: $\mathbf{S}=\sum\limits_{i}\mathbf{s}_{i},
\mathbf{L}=\sum\limits_{i}\mathbf{l}_{i},\mathbf{T}=\sum\limits_{i}
\left(\mathbf{s}_{i}-3\mathbf{r}_{i}\left(
\mathbf{r}_{i}\mathbf{s}_{i}\right) /r_{i}^{2}\right)$ with
$\mathbf{s}_{i}$ and $\mathbf{l}_{i}$ being the spin and  orbital
moments for $i$-th electron, $\mathbf{r}_{i}$ the coordinate operators.

A simple derivation of the XMCD sum rules in the independent
electron approximation was presented in Ref.~\cite{rehr}. This
derivation is based on a purely band picture of electron states in
solids. The opposite case of strongly localized electrons which
are characterized by atomic states with well-defined term and
multiplet structure has been considered in Refs.~\cite{carra,laan}
(a somewhat more simple derivation with the use of second
quantization formalism for the atomic states was presented in the
paper \cite{carra1}). At the same time, many interesting systems
such as magnetite \cite{huang} should be considered as strongly
correlated systems demonstrating simultaneously both itinerant and
localized features of ``magnetic'' electrons \cite{vons,FeNi}.
Actually, the sum rules are widely exploited by experimentalists
for such systems as well. In this work we present a formal
justification of this simple way to proceed. At the same time, we
discuss separately contributions to the sum rules from coherent
and incoherent parts of the electron spectral density. We
demonstrate that a proper account of the incoherent
(nonquasiparticle) contributions is necessary for consistent
treatment of the XMCD spectra of strongly correlated systems.

The spectral intensity $\mu(\omega)$ of the X-ray absorption and
emission spectra (XAS and XES, respectively) is determined in the
dipole approximation by the imaginary part of the Green's function
\cite{mahan,zubarev}
\begin{eqnarray}
\left\langle \left\langle \mathbf{p\cdot e}^{\ast }|\mathbf{p\cdot e}%
\right\rangle \right\rangle _{\omega } =e_{\alpha }^{\ast
}e_{\beta
}G_{\alpha \beta }\left( \omega \right) \\
G_{\alpha \beta }\left( \omega \right)  =\left\langle \left\langle
p_{\alpha }|p_{\beta }\right\rangle \right\rangle _{\omega
}=-i\int\limits_{0}^{\infty }dte^{i\omega t}\left\langle \left[
p_{\alpha }\left( t\right) ,p_{\beta }\right] \right\rangle
\label{zub}
\end{eqnarray}
where $\mathbf{e}$ is the photon polarization vector and
$\mathbf{p}$ is the momentum operator, $\alpha ,\beta $ are the
Cartesian indices and the brackets stand for the Gibbs average in
the initial state. For the case of XAS only the part of this
operator works which corresponds to the transitions from core
states $\left| a\right\rangle $ (with the annihilation operators
$b_{a}$) to the conduction electron states $\left| \lambda
\right\rangle $ created by operator $c_{\lambda }^{\dagger }$%
\begin{equation}
p_{\alpha }^{\left( +\right) }=\sum\limits_{a\lambda }\left\langle
\lambda \right| p_{\alpha }\left| a\right\rangle c_{\lambda
}^{\dagger }b_{a}, \label{operator}
\end{equation}
and only the Hermitian conjugated part $p_{\alpha }^{\left(
-\right) }$ works for the XES. For the case of L$_2$(L$_3$)
spectra $a$ labels total moment projection for $2p_{1/2}$
$(2p_{3/2})$ states, correspondingly, and $\lambda$ labels spin
projection and orbital indices for $3d$-electrons. Transitions to
$s$-states, which are also allowed in the dipole approximation,
are irrelevant for magnetism (in particular, they practically do
not contribute to XMCD) and therefore will be neglected further.

It is important that Eqs.(\ref{zub}), (\ref{operator}) are
formally exact irrespective to the degree of localization or
delocalization of $d$-electrons. The use of the atomic
representation for the states $\left| \lambda \right\rangle$
allows to obtain explicit expressions for the matrix elements
$\left\langle \lambda \right| p_{\alpha }\left| a\right\rangle$ in
terms of $3nj$-symbols, fractional parentage coefficients and
irreducible matrix elements \cite{carra,carra1,laan}. However, for
the derivation of the sum rules such a concretization is not
necessary and using the $\lambda$-representation yields a simpler
way to proceed. Actually, total intensities of the XAS for
different photon polarizations which are needed to obtain the XMCD
sum rules are determined by the integrals of the spectral density
in the infinite limits. Due to the Kramers-Kronig relations, the
latter are equal to
\begin{equation}
-\int\limits_{-\infty }^{\infty }\frac{d\omega}{\pi} {\rm Im}G_{\alpha \beta
}\left( \omega \right) = \lim_{\omega \rightarrow \infty
}\left[ \omega {\rm Re}G_{\alpha \beta
}\left( \omega \right) \right] =\left\langle \left[ p_{\alpha },p_{\beta }%
\right] \right\rangle   \label{limit}
\end{equation}
Here we have used the equations of motion for the Green's
functions \cite{zubarev}. On substituting Eq.(\ref{operator}) into
the right-hand side of Eq.(\ref{limit}) one can see that the
commutator contains just the one-particle density matrices which
are expressed in terms of the corresponding anticommutator Green's
functions:
\begin{equation}
\rho_{\lambda' \lambda}=\langle c_{\lambda }^{\dagger }c_{\lambda
^{\prime }}\rangle=-\int\limits_{-\infty }^{\infty }\frac{dE}{\pi}
f(E){\rm Im} \langle \langle c_{\lambda ^{\prime }}|c_{\lambda
}^{\dagger }\rangle \rangle_{E} \label{rho}
\end{equation}
where $f(E)$ is the Fermi function; similarly we introduce the
core density matrix
\begin{equation}
\rho^{\rm core}_{a'a}=\left\langle b_{a}^{\dagger }b_{a^{\prime
}}\right\rangle \label{rho1}
\end{equation}
Assuming that in the initial state the core electron states are
completely occupied, i.e., $\rho^{\rm core}_{aa'}=
\delta_{aa^{\prime}}$, one can receive the expression for the
total spectral weight (integrated intensity) of L$_{2,3}$ spectra:
\begin{eqnarray}
\overline{\mu}_j =2\pi \sum\limits_{m_{j}m\sigma m^{\prime }\sigma
^{\prime }}\langle jm_{j}|lm\frac{1}{2}\sigma
\rangle \langle lm\frac{1}{2}\sigma | \mathbf{p\cdot e}^{\ast} \\
\nonumber
\times (\widehat{\mathbf{1}}-\widehat{\mathbf{\rho }}) \mathbf{p\cdot e}%
| lm^{\prime }\frac{1}{2}\sigma ^{\prime }\rangle \langle
lm^{\prime }\frac{1}{2}\sigma ^{\prime }|jm_{j}\rangle
\label{rehr}
\end{eqnarray}
which differs from Eq.(2) of Ref.~\cite{rehr} only by the
replacement of one-electron expression for the density matrix
$\rho_{mm'}$ by the exact one. Here $l=2$, $j=1/2$ (3/2) for L$_2$
(L$_3$) spectra, $m,m^{\prime }$ and $\sigma ,\sigma ^{\prime }$
are the orbital and spin projections of $d$-electrons,
respectively, $m_j$ is the total moment projection for the core
states. It is worthwhile to stress that using the orbital indices
for itinerant electrons does not mean any approximation: for any
particular method of band structure calculations it is always
possible to re-expand the Wannier functions at a given site into
the spherical harmonics. Thus the density matrix $\rho_{mm'}$ is
in general a linear combination of the band structure occupation
numbers with a proper symmetry. Since the total spin and orbital
moment as well as the spin dipole operator are one-particle
operators, their averages are completely determined by the density
matrix. Further use of the Wigner-Eckart theorem to extract
angular dependences of the matrix elements and transformations of
the arising 3$j$-symbol products repeat the derivation in
Ref.~\cite{rehr}. Therefore the XMCD sum rules (\ref{sumrules})
are formally valid without any restrictions.

On the other hand, the values of $\left\langle S_{z}\right\rangle$
and $\left\langle L_{z}\right\rangle$ obtained in band
calculations can violate the sum rules. The calculation of the
one-particle density matrix for strongly correlated systems
remains a quite nontrivial problem. Generally, this quantity
contains both coherent (quasiparticle) contributions which are
formally connected with the poles of the Green's function and
incoherent (nonquasiparticle) ones which are formally connected
with the branch cuts \cite{nozieres}. In a number of cases (for
example, for strongly correlated metals in the vicinity of the
Mott transition or for doped Mott insulators \cite{GKKR}) the
quasiparticle spectral density is much narrower than the
incoherent one. One should be therefore careful to avoid a
confusion of the incoherent part with a spectrum background.

In practice, the use of the XMCD sum rules for strongly correlated
systems is not a completely well-defined procedure since it is
impossible to integrate the spectral density in infinite limits.
Atomic multiplet structure leads sometimes to much broader
distribution of the spectral density in comparison with a standard
band picture. This fact has been recently demonstrated
\cite{wessely} for the case of (La,Sr)MnO$_3$ where the
configuration-interaction calculations for Mn$^{3+}$ ion gave the
spectral density with the width of order of 6$\div$7 eV, in
comparison with the value 2$\div$3 eV within the local density
approximation (LDA) or LDA+U. At the same time, the energy
separation between L$_2$ and L$_3$ spectra varies from 6$\div$8 eV
for the light $3d$ metals (Ti, V) to 15$\div$20 eV for heavy ones
(Co, Ni, and Cu). This means that for strongly correlated systems
L$_2$ and L$_3$ spectra can overlap appreciably. The energy
distribution of the atomic multiplet structure can be comparable
with the spin-orbit splitting of the relevant core levels also for
rare earth systems which leads to some problems with the practical
applications of the XMCD sum rules in the latter case
\cite{teramura}.

It is a common practice to interpret the core-level spectra for
isolated atoms in terms of many-electron multiplet picture rather
than one-electron quantum numbers $|\lambda \rangle$ used above.
These two approaches can be related via the representation of
one-electron operators in terms of the Hubbard X-operators
\cite{hubbardIV} $X^{\Gamma,\Gamma'}=|\Gamma \rangle\langle
\Gamma'|$ where $\Gamma=\{ nLSM \Sigma \}$ labels atomic
configuration $n$, term $LS$ and moment projections $M,\Sigma$:
\begin{eqnarray}
c_{lm\sigma }^{\dagger }=\sum\limits_{n\Gamma _{n}\Gamma _{n-1}}\sqrt{n}%
G_{\Gamma _{n-1}}^{\Gamma _{n}}\langle L_{n}M_{n}|L_{n-1}M_{n-1}lm \rangle \nonumber \\
\times \langle S_{n}\ \Sigma_{n}|S_{n-1} \Sigma_{n-1}\frac{1}{2}%
\sigma \rangle X^{\Gamma _{n},\Gamma _{n-1}}  \label{xoper}
\end{eqnarray}
where $G_{\Gamma _{n-1}}^{\Gamma _{n}}$ are the fractional
parentage coefficients (see Refs.~\cite{judd,ii}).
Eq.(\ref{xoper}) enables one to reproduce the sum rules in the
form yielding a detailed information concerning term and multiplet
structure \cite{carra,carra1,laan}.

At the same time, for a solid the atomic description can be
inappropriate. In particular, ferromagnetism itself is an
essentially band phenomenon. Moreover, it cannot be properly
described in the simplest Hubbard-I approximation \cite{hubbardI}
which assumes a formation of individual Hubbard subbands from
separate transitions between the atomic levels. In particular, for
a narrow-band ferromagnet we cannot satisfy the sum rules
(kinematic relations) for the X-operators in this approximation.

To demonstrate this we consider a simple model of a narrow-band
itinerant electron ferromagnet with orbital degrees of freedom
which generalizes the standard infinite-$U$ Hubbard model:
\begin{equation}
\mathcal{H}=\sum_{\mathbf{k}m \sigma
}t_{\mathbf{k}m}X_{-\mathbf{k}}^{0,\sigma m
}X_{\mathbf{k}}^{\sigma m, 0} \label{eq:HHM}
\end{equation}
where $t_{\mathbf{k}m}$ is the orbital-dependent band energy,
$X_{\mathbf{k}}^{\alpha, \beta }$ is the Fourier transform of the
Hubbard operators and 0 labels the hole state at a site.

For this model the exact sum rules should be satisfied for arbitrary $m,\sigma$:
\begin{eqnarray}
\delta \equiv n_0 &=& \langle X^{0,0}\rangle =\langle
X_i^{0,\sigma m }X_i^{\sigma m, 0}\rangle
=\sum\limits_{\mathbf{k}}\langle X_{-\mathbf{k}}^{0, \sigma
m}X_{\mathbf{k}}^{\sigma m, 0}\rangle
\nonumber \\
&=& - \sum_{\mathbf{k}}\int\limits_{-\infty }^{+\infty }
\frac{dE}{\pi} f(E) {\rm Im} \langle \langle
X_{\mathbf{k}}^{\sigma m, 0}|X_{-\mathbf{k}}^{0, \sigma m} \rangle
\rangle_{E}. \label{eq:n0}
\end{eqnarray}
where $\delta$ is the concentration of current carriers (holes). The Hubbard-I
approximation for this model reads
\begin{equation}
\langle \langle X_{\mathbf{k}}^{\sigma m, 0}|X_{-\mathbf{k}}^{0,
\sigma m} \rangle \rangle_{E}=[E-t_{\mathbf{k}m}(N_0+N_{\sigma
m})]^{-1}. \label{eq:GF:0}
\end{equation}
According to this expression, the quasiparticle pole for $\sigma
=\downarrow$ corresponds to a narrowed band and lies (for the
saturated ferromagnet case) above the Fermi level of the holes
which obviously violates the sum rule (\ref{eq:n0}). In fact,
these sum rules are satisfied only due to incoherent
(nonquasiparticle) states which are present below the Fermi level
for the Hubbard model \cite{Irkhin1983}. Similar to these papers
one obtains for the minority-spin Green's functions in the leading
order in small parameter $\delta$
\begin{eqnarray}
-\frac{1}{\pi} {\rm Im} \langle \langle X_{\mathbf{k}}^{\downarrow
m,0}|X_{-\mathbf{k}}^{0, \downarrow m} \rangle \rangle_{E} \nonumber \\
=\sum_{\mathbf{q}m'}f(E_{\mathbf{k-q}\uparrow m'})
\delta(E-E_{\mathbf{k-q}\uparrow m'} +\omega _{\mathbf{q}mm'}).
\label{inc)}
\end{eqnarray}
where $E_{\mathbf{k} \uparrow m}=t_{\mathbf{k}m}(N_0+N_{\uparrow
m})$, $\omega _{\mathbf{q}mm'}$ are the frequencies of the
corresponding spin-flip transitions. Thus we have 100\% incoherent
contribution to the spectral density below the Fermi level. As the
hole concentration increases, the ferromagnetic state becomes
non-saturated and a narrow quasiparticle minority-spin band occurs
\cite{iz04}. However, the main spectral density for this spin
projection is still due to the nonquasiparticle contribution. This
example demonstrates that, despite the XMCD sum rules for strongly
correlated ferromagnets have formally a standard ``one-electron''
form, the energy distribution of the spectral weight can be
drastically different from usual band picture.

The theoretical consideration \cite{EPJ2005} shows that the
nonquasiparticle contributions to XES and XAS should be clearly
observed. Moreover, such contributions should be considerably
enhanced by the interaction with the core hole \cite{EPJ2005}. As
well as for XES and XAS in general, the main nonquasiparticle
effects in the XMCD are connected with the occurrence of the
incoherent spectral density in the energy gap for the
half-metallic ferromagnets \cite{ufn}.
These states arise either only below
the Fermi energy (for the majority-gap half-metallic ferromagnets
such as magnetite) or only above it (for the minority-gap
half-metallic ferromagnets such as Heusler alloys or CrO$_2$);
therefore they can be studied by XES and XAS methods,
correspondingly. Whereas a standard band theory predicts 100\%
polarization for the conduction electron (or hole) states, the
depolarization due to the nonquasiparticle states can be very
strong (for example, in the infinite-$U$ Hubbard model limit there
is no polarization at all \cite{Irkhin1983}). At the same time,
{\it ab initio} calculations of the correlation effects for the
half-metallic Heusler alloy NiMnSb give a rather small spectral
weight of the nonquasiparticle states (about 4\%) \cite{lulu} and
therefore it would be preferable to investigate them for the
half-metallic ferromagnets with more strong correlations such
as Fe$_3$O$_4$ (by XES) and CrO$_2$ (by XAS).

The research described was supported in part by Grant No.
747.2003.2 from the Russian Basic Research Foundation (Support of
Scientific Schools), by Russian Science Support Foundation
and by the Netherlands Organization for Scientific Research
(Grant NWO 047.016.005).

\end{document}